\documentstyle[epsf,aps,preprint,floats,tighten]{revtex}

\begin{document}
		
\preprint{\vbox{\hbox{UM-TH-96-10}
		\hbox{hep-ph/9607448}
		}
}

\title{Goldstino Decoupling in Spontaneously Broken Supergravity Theories}

\author{Tony Gherghetta\footnote{tgher@umich.edu}}
\address{Department of Physics, University of Michigan,
	Ann Arbor, Michigan~~48109-1120}

\date{July, 1996}

\maketitle

\begin{abstract}
If the gravitino is sufficiently light and stable it will behave as an 
effective massless neutrino species at the time of nucleosynthesis. 
Depending on the temperature at which it decouples from the thermal bath 
in the early universe, the gravitino 
mass will be bounded by the primordial $^4$He abundance. Assuming a 
conservative estimate that the number of neutrino families, $N_\nu < 3.6$,
superlight gravitinos with a mass $m_{3/2} \lesssim 10^{-6}$eV are ruled out.
This bound is weaker than previous estimates because the Goldstino
annihilation cross section was overestimated. 
\end{abstract}

\newpage

\section{Introduction}

If supersymmetry is to describe the real world then it must be broken
at some scale $\Lambda$ and give rise to supersymmetry breaking mass 
splittings not greater than $\tilde{m}\sim {\cal O}$(TeV). The signal 
for local supersymmetry breaking occurs when the superpartner of the 
graviton, the gravitino, receives a mass via the superHiggs mechanism. 
There is an upper 
bound of $\cal O$(TeV) on the gravitino mass which comes from requiring that 
supersymmetry provides the solution to the naturalness problem. 
However, the gravitino mass is not necessarily constrained to be 
${\cal O}$(TeV) because the gravitino only interacts gravitationally
and so is weakly coupled (compared to the gauge forces).
This is not the case for the 
superpartners of the particles in the standard model which experience gauge 
forces. Their soft masses are constrained from experiment to be in the 
range ${\cal O}(10^2-10^3)$GeV.

The relationship between the gravitino mass and the soft masses depends
on the messenger sector. 
The messenger sector is responsible for communicating the spontaneous
breakdown of supersymmetry from a hidden sector, where the supersymmetry 
breaking dynamics occurs, to the observable sector.
Normally in hidden sector scenarios of spontaneous supersymmetry breaking, 
the messenger sector is gravitational and gives rise to a gravitino mass 
$m_{3/2} \simeq \tilde{m}$. A light gravitino is precluded because it would 
also mean light gauginos, sleptons and squarks. This typically assumes
generic choices of the Kahler potential, ${\cal G}(z,z^\ast)$ and
the gauge kinetic function $f_{ab}(z)$ where $z$ is the hidden sector 
chiral superfield. However, in no-scale supergravity models particular 
choices of ${\cal G}(z,z^\ast)$ and $f_{ab}(z)$ make it possible for 
the gravitino mass $m_{3/2}\ll \tilde{m}$ \cite{ellis}.

A hierarchy between $m_{3/2}$ and $\tilde{m}$ is also possible if 
the messenger sector responsible for communicating supersymmetry breaking 
to the visible sector is not gravitational. 
Recently there has been a renewed interest in dynamical 
supersymmetry breaking models \cite{dnns,dn,dns} where gauge forces 
communicate the breakdown of supersymmetry to the squarks, sleptons and 
gauginos. In these models the soft masses $(\tilde{m})$ are generated by 
radiative corrections and are proportional to the gauge couplings squared.
The gauge sparticles will still receive soft masses from gravitational 
interactions, but these contributions are much smaller than the 
contributions from the gauge-mediated messenger sector.
Since the gravitinos do not carry gauge quantum numbers they do not receive 
any mass from the gauge messenger sector but still receive mass from 
gravitational interactions.  Consequently, in these models the gravitino 
is naturally light ($\ll \cal O$(TeV)) and becomes the lightest 
supersymmetric particle (LSP). This can have interesting
cosmological implications, such as contributing to the dark matter 
component of the universe \cite{bmy}. 

Given the possibility of non-minimal kinetic terms or a non-gravitational 
messenger sector one can ask how light can the gravitino mass be?  A naive 
estimate of this mass comes
from assuming that the supersymmetry breaking scale $\Lambda \sim 10^2$GeV, 
which gives $m_{3/2}\sim \Lambda^2/M_{Pl} \sim 10^{-6}$eV where 
$M_{Pl}$ is the Planck mass. If the gravitino were this light 
then it would essentially behave as a massless neutrino species during
nucleosynthesis. This is because at energies $E\gg m_{3/2}$ the longitudinal
component of the gravitino dominates during interactions, which is just a 
statement of the Equivalence Theorem \cite{cddfg}. The longitudinal 
components of the gravitino are the helicity $\pm 1/2$ modes which come 
from absorbing the spin-1/2 Goldstone fermion (or Goldstino) during the 
spontaneous breakdown of supersymmetry.
The Goldstino component of the gravitino couples with a strength proportional
to $\kappa/m_{3/2}$, where $\kappa = 1/M_{pl}$. If the gravitino mass is 
light this coupling can be much stronger than naively expected from
the gravitational force. 
This means that as the temperature of the universe cools, the Goldstino 
will remain in thermal contact with the heat bath longer because of its 
enhanced coupling. Since the expansion rate of the universe depends on the 
number of degrees of freedom, the universe will expand faster with 
Goldstinos present at temperatures $T\lesssim {\cal O}$(100 MeV) and 
cause the neutrinos to decouple earlier. This will affect the production of 
neutrons (via the charged current weak interactions) by causing more
neutrons to survive and ultimately increases the primordial $^4$He abundance. 
Thus the observed primordial $^4$He abundance sets a lower bound on the 
gravitino mass, which in turn is related to how late the Goldstino 
decouples from the thermal bath.

A lower bound on the gravitino mass based on the nucleosynthesis argument
was originally discussed by Fayet
\cite{fayet}, who considered the effect of the Goldstino together with a 
light photino on the effective number of massless neutrino species. The 
bound derived by Fayet $(m_{3/2}\gtrsim 10^{-2}\,$eV) is not effective 
anymore because the photino is now
believed to be of order the electroweak scale and the gravitino 
(if it is light enough) can only thermally interact with leptons and photons 
below the quark-hadron phase transition. More recently Moroi et al 
\cite{moroi} have also considered this bound and obtain an estimate 
$m_{3/2}\gtrsim 10^{-4}$eV. However the cross section used by them to 
recalculate the Fayet 
bound is only valid for $T\gg {\tilde m}, m_{3/2}$. In this work we will 
calculate the Goldstino decoupling temperature 
using the cross section for Goldstino annihilation in the limit 
$\tilde{m} \gg T \gg m_{3/2}$, where we assume that the supersymmetry breaking
mechanism gives rise to a superlight gravitino. The gravitino mass bound will
be shown to be much weaker than has previously been estimated.

The outline of this paper is as follows. We begin in Sec.2 with a 
discussion of the effective interaction Lagrangian for the Goldstino. This
Lagrangian is used to calculate the Goldstino annihilation cross section 
into leptons and photons. In Sec.3 the thermally averaged Goldstino
annihilation rate is obtained. This will be used to calculate the
Goldstino decoupling temperature in Sec.4. A lower bound on the gravitino 
mass and the low energy supersymmetry breaking scale will be derived from
the Goldstino contribution to the effective number of massless neutrino 
species. Final comments and the conclusion will be presented in Sec.5.

\section{Effective Goldstino Lagrangian}

The effective Lagrangian for the gravitino is obtained from the N=1 
supergravity theory \cite{cremmer}, irrespective of the mechanism 
employed in the messenger sector to communicate supersymmetry breaking 
to the observable sector. Denoting the chiral matter supermultiplets by 
$(\phi,\chi)$ and the gauge vector supermultiplets by 
$(\lambda,A_\mu)$ the supergravity Lagrangian for the interaction of the
gravitino with the vector and chiral multiplets is given by
\begin{eqnarray}
	\label{sgl}
	{\cal L} &=&{\kappa\over 4}~\bar{\lambda} \gamma^\mu 
	\sigma^{\alpha\beta}\Psi_\mu~F_{\alpha\beta} 
	+ {\kappa\over 4}~h_{\mu\nu}\left[\eta^{\alpha\beta}(\eta^{\mu\rho}
	\eta^{\nu\sigma}+\eta^{\mu\sigma}\eta^{\nu\rho})~F_{\rho\alpha}~
	F_{\sigma\beta} - {1\over 2} \eta^{\mu\nu}~F_{\alpha\beta}~
	F^{\alpha\beta}\right] 
	\nonumber \\
	&+& \left[{\kappa\over\sqrt{2}} \bar{\Psi}_\mu
	\gamma^\nu \gamma^\mu \Psi_{iR} D_\nu\phi^{i\ast}
	-{\kappa^2\over 16} \bar{\Psi}_L^i\gamma_\sigma \Psi_{Li}
	(i\epsilon^{\mu\nu\rho\sigma} \bar{\Psi}_\mu \gamma_\nu \Psi_\rho
	- \bar{\Psi}_\mu\gamma_5 \gamma^\sigma \Psi^\mu) + h.c. \right] 
	\nonumber\\
	&+&i{\kappa\over 8} h_{\mu\nu}\left[\bar{\Psi}_i(\gamma^\nu
	\partial^\mu
	+\gamma^\mu\partial^\nu)\Psi_i - (\partial^\mu\bar{\Psi}_i\gamma^\nu
	+\partial^\nu\bar{\Psi}_i\gamma^\mu)\Psi_i\right] 
	\nonumber \\
	&-& {\kappa\over 4} m_{3/2}~(\eta^{\lambda\mu}\eta^{\rho\nu}
	+\eta^{\lambda\nu}\eta^{\rho\mu}-\eta^{\mu\nu}\eta^{\lambda\rho})
	\bar{\Psi}_\mu\Psi_\nu ~h_{\lambda\rho}
	-i {\kappa\over 4} \epsilon^{\mu\sigma\nu(\lambda} \bar{\Psi}_\mu
	~\gamma_5~[\gamma_\sigma,\sigma^{\rho)\tau}]_+~\Psi_\nu
	~\partial_\tau h_{\lambda\rho} 
	\nonumber \\
	&+& {\kappa\over 4}
	\left[\epsilon^{\mu\sigma\nu(\lambda} \bar{\Psi}_\mu\gamma_5
	\gamma^{\rho)}\partial_\sigma\Psi_\nu~h_{\lambda\rho} -
	\epsilon^{\mu\sigma\nu(\lambda} \partial_\sigma\bar{\Psi}_\mu
	\gamma_5 \gamma^{\rho)} \Psi_\nu~h_{\lambda\rho}\right] 
	\nonumber \\
	&+& {\kappa\over 4} \sum_j~c_j~F_{\mu\nu}F^{\mu\nu}~S_j 
	+ i{\kappa\over 2} m_{3/2}~\sum_j d_j~\bar{\Psi}_\mu\sigma^{\mu\nu}
	\Psi_\nu~S_j 
	\nonumber \\
	&+& {\kappa\over 8} \sum_j~c_j~\epsilon^{\mu\nu\rho\sigma}
	F_{\mu\nu}F_{\rho\sigma}~P_j 
	+ i{\kappa\over 4}\sum_j d_j~\epsilon^{\mu\nu\rho\sigma}
	\bar{\Psi}_\mu \gamma_\nu \Psi_\rho \partial_\sigma P_j
	-{\kappa\over 2}\sum_j~d_j \bar{\Psi}_{Li}\gamma^\mu\Psi_{Li}
	\partial_\mu P_j
\end{eqnarray}
where $\sigma^{\mu\nu}={i\over 2}[\gamma^\mu,\gamma^\nu]$, $\Psi_\mu$ 
denotes the gravitino and $\Psi_i=(\chi_i,\bar{\chi}_i)^T$ is a Majorana 
spinor. The symmetric tensor 
$h_{\mu\nu}$ denotes the graviton where we have written the metric
tensor as $g_{\mu\nu}(x)=\eta_{\mu\nu}+\kappa h_{\mu\nu}(x)$. 
The hidden sector scalar 
components are defined as $S_j=1/\sqrt{2}(\phi_j+\phi_j^\ast)$ and
$P_j=1/(\sqrt{2}i)(\phi_j-\phi_j^\ast)$, and the constants $c_j,d_j$
satisfy
\begin{equation}
\label{cddef}
	\sum_j c_j d_j = {m_{\tilde g}\over m_{3/2}}
\end{equation}
where $m_{\tilde{g}}$ is the gaugino mass, $\langle f_{ab}\rangle =
\langle f\rangle \delta_{ab}$ and we have adopted the normalising 
conventions used in Ref.\cite{ulbr}. In the Lagrangian (\ref{sgl})
we have neglected all terms which are not relevant for the calculation
of the gravitino annihilation cross section.

The Lagrangian (\ref{sgl}) may be used to calculate scattering 
processes involving all the helicity
components of the massive gravitino. However, in the limit that the
energy scale of the gravitino $E\gg m_{3/2}$,
the longitudinal component of the gravitino will dominate and the gravitino
effectively behaves as a spin-1/2 Goldstino
\begin{equation}
	\label{gold}
	\Psi_\mu \sim i\sqrt{2\over3} {1\over m_{3/2}} \partial_\mu 
	\widetilde{G}
\end{equation}
where $\widetilde G$ denotes the Goldstino and the factor $\sqrt{2\over3}$ 
is a Clebsch-Gordon coefficient. This is just a statement of the Equivalence 
Theorem and is analogous to longitudinal W bosons in the standard electroweak
model behaving as Nambu-Goldstone bosons in the high energy limit. If we 
substitute (\ref{gold}) into the supergravity Lagrangian (\ref{sgl}) 
and make use of the fact that for an on-shell gravitino 
$\gamma^\mu \Psi_\mu =0$ then we obtain an effective Lagrangian for the 
interaction of the Goldstino with gauge and matter multiplets.

\begin{figure}
	\centering
	\epsfxsize=5.0in
	\hspace*{0in}
	\epsffile{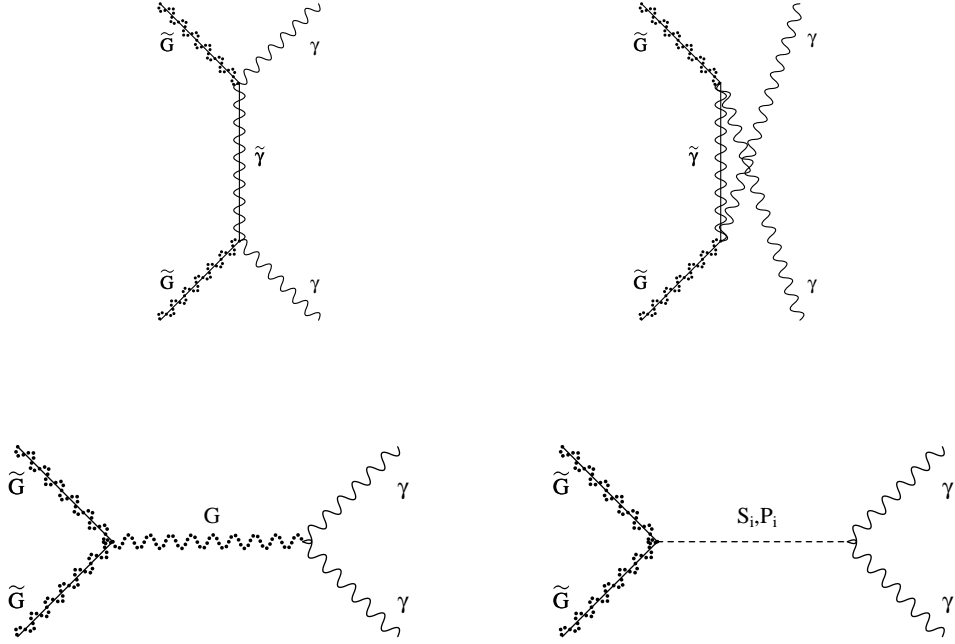}
	\caption{\it Feynman diagrams for the scattering process
	$\widetilde{G}\widetilde{G}\rightarrow \gamma\gamma$. }
\label{fig:FDGGgg}
\end{figure}

\begin{figure}
	\centering
	\epsfxsize=5.0in
	\hspace*{0in}
	\epsffile{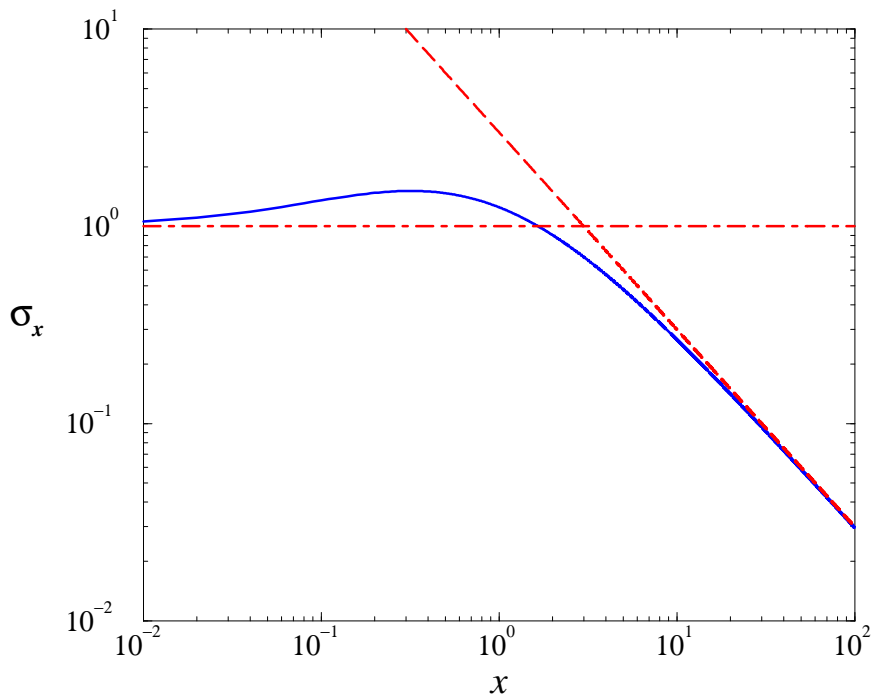}
	\caption{\it Comparison of the various limits for the cross section
	$\sigma({\widetilde G}{\widetilde G}\rightarrow \gamma\gamma)$ where
	only the quantity in square brackets $(\equiv\sigma_x)$ in 
	Eq.(\ref{GGggcs}) is plotted.
	The solid curve represents the exact expression for $\sigma_x$,while 
	the dot-dashed line depicts $\sigma_x$ in the Equivalence Theorem 
	limit $(\protect\sqrt{s}\gg\tilde{m})$ and the dashed line 
	represents $\sigma_x$ 
	in the intermediate limit $(\tilde{m}\gg\protect\sqrt{s})$.}
\label{fig:GGgg}
\end{figure}

The annihilation channels that will interest us are Goldstino scattering
to leptons and photons. We will assume that all other particles have 
decoupled and are not important for the calculation of the Goldstino 
decoupling temperature. Consider first Goldstino annihilation into photons. 
There are three different contributions to this amplitude which are 
depicted in Fig.~\ref{fig:FDGGgg}. 
The first contribution is due to the $t$ and $u$-channel exchange of the 
photino, where for simplicity we ignore neutralino mixing. In addition there
are $s$-channel annihilation diagrams which result from graviton exchange and 
hidden sector scalar exchange. These diagrams are important for the 
cancellation of the leading order energy dependence in the Equivalence 
Theorem limit \cite{br}. The Lagrangian relevant for these processes 
is given by
\begin{eqnarray}
	\label{GGgglag}
	{\cal L} &=& {1\over 2\sqrt{6}} {\kappa\over m_{3/2}} \partial_\mu
	\bar{\lambda} \gamma^\mu [\gamma^\nu \gamma^\rho] 
	\widetilde{G} \partial_\nu A_\rho 	+ {\kappa\over 4}~h_{\mu\nu}
	\left[\eta^{\alpha\beta}(\eta^{\mu\rho}
	\eta^{\nu\sigma}+\eta^{\mu\sigma}\eta^{\nu\rho})~F_{\rho\alpha}~
	F_{\sigma\beta} - {1\over 2} \eta^{\mu\nu}~F_{\alpha\beta}~
	F^{\alpha\beta}\right] 
	\nonumber \\
	&-&{1\over 6}~{\kappa\over m_{3/2}} ~(\eta^{\lambda\mu}
	\eta^{\rho\nu}+\eta^{\lambda\nu}\eta^{\rho\mu}-\eta^{\mu\nu}
	\eta^{\lambda\rho})~\partial_\mu\overline{\widetilde{G}}~
	\partial_\nu\widetilde{G}~h_{\lambda\rho} 
	\nonumber \\
	&+&{i\over 6}~{\kappa\over m_{3/2}^2}~\epsilon^{\mu\sigma\nu(\lambda}
	~\partial_\mu\overline{\widetilde{G}}~\gamma_5~[\gamma_\sigma,
	\sigma^{\rho)
	\tau}]_+~\partial_\nu\widetilde{G}~\partial_\tau h_{\lambda\rho}
	\nonumber \\
	&+& {\kappa\over 2} \sum_i c_i~(\partial_\nu A_\mu
	\partial^\nu A^\mu - \partial^\nu A_\mu \partial^\mu A_\nu) S_i
	+ {1\over 3} {\kappa\over m_{3/2}} \sum_i d_i~\partial_\mu 
	\overline{\widetilde{G}} \partial^\mu \widetilde{G} S_i	
	\nonumber \\
	&+& {\kappa\over 2} \sum_i c_i~\epsilon^{\mu\nu\rho\sigma} 
	\partial_\mu A_\nu \partial_\rho A_\sigma P_i
	- {1\over 6} {\kappa\over m_{3/2}^2} \sum_i d_i~\partial_\mu 
	\overline{\widetilde{G}} \gamma^\nu\gamma_5 \partial^\mu 
	\widetilde{G} \,\partial_\nu P_i.
\end{eqnarray}
Assuming that $\sqrt{s} \gg m_{3/2}$ the Goldstino
annihilation cross section into photons is calculated using the
above Lagrangian (\ref{GGgglag}) to be
\begin{equation}
	\label{GGggcs}
	\sigma(\widetilde{G}\widetilde{G}\rightarrow \gamma\gamma)
	={1\over 1728\pi} {\kappa^4\over m_{3/2}^4}
	m_{\tilde\gamma}^4 s\left[1+{6x(1-2x-4x^2)\over 1+x} 
	+{6x(-x+4x^2+8x^3)\over 1+2x} \log (1+{1\over x})\right].
\end{equation}
where $x=m_{\tilde\gamma}^2/s$. This agrees with the result quoted in Ref.
\cite{br}. In the limit that $\sqrt{s} \gg m_{\tilde\gamma}$ the cross 
section is in agreement with the Equivalence Theorem, namely
\begin{equation}
	\label{GGggcset}
	\sigma(\widetilde{G}\widetilde{G}\rightarrow \gamma\gamma)
	\simeq {1\over 1728\pi} {\kappa^4\over m_{3/2}^4}
	m_{\tilde\gamma}^4 s\simeq f^4 {\kappa^4 s\over 1728\pi} 
\end{equation}
where $f$ generically denotes the photino ``Yukawa'' coupling.
On the other hand in the intermediate limit $m_{\tilde{\gamma}} 
\gg \sqrt{s}$ the cross section (\ref{GGggcs}) becomes
\begin{equation}
	\label{GGggcsil}
	\sigma(\widetilde{G}\widetilde{G}\rightarrow \gamma\gamma)
	={1\over 576\pi} {\kappa^4\over m_{3/2}^4}
	m_{\tilde\gamma}^2 s^2
\end{equation}
which has a different dependence on the energy scale $\sqrt{s}$ than in
(\ref{GGggcset}).
This will be the limit that is needed to accurately calculate the Goldstino
decoupling temperature. In Fig.~\ref{fig:GGgg} these various limits are 
compared with the exact result. It is clear from the figure that the cross
section (\ref{GGggcset}) is a poor approximation in the limit 
$m_{\tilde{\gamma}}\gg\sqrt{s}$.

Notice also that the cross section (\ref{GGggcset}) is proportional to $E^2$.
This is due to the nonrenormalisibility of the supergravity Lagrangian and
can lead to unitarity violation above a critical energy \cite{ulbr},
\begin{equation}
	\label{ecr}
	E_{cr}=\sqrt{288\pi} {m_{3/2}\over \kappa m_{\tilde{g}}}.
\end{equation}
If the critical energy is taken to be the Planck scale, then it is difficult
to reconcile a light gravitino mass with (\ref{ecr}). However it was 
pointed out in 
Ref.\cite{spm} that one can interpret the critical energy as the scale of 
new physics at which the gaugino mass, $m_{\tilde{g}}$ becomes effective. 
This scale need not necessarily be anywhere near the Planck scale. For a
gravitino mass $m_{3/2}\sim{\cal O}(10^{-6}{\rm eV})$ the critical energy 
turns out to be $E_{cr}\sim 3\,$TeV. This unitarity limit is safely above
the energy scales that we will be concerned with, which are
below the quark-hadron phase transition $(\sim {\cal O}(100\,{\rm MeV}))$.

\begin{figure}
	\centering
	\epsfxsize=5.0in
	\hspace*{0in}
	\epsffile{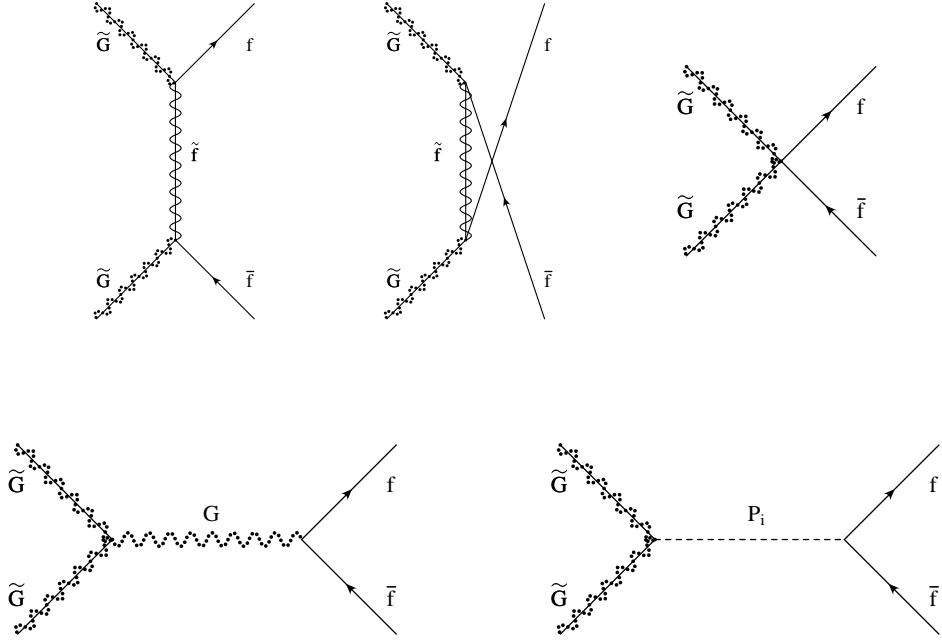}
	\caption{\it Feynman diagrams for the scattering process
	$\widetilde{G}\widetilde{G}\rightarrow f\bar{f}$. }
\label{fig:FDGGff}
\end{figure}

\begin{figure}[t]
	\centering
	\epsfxsize=5.0in
	\hspace*{0in}
	\epsffile{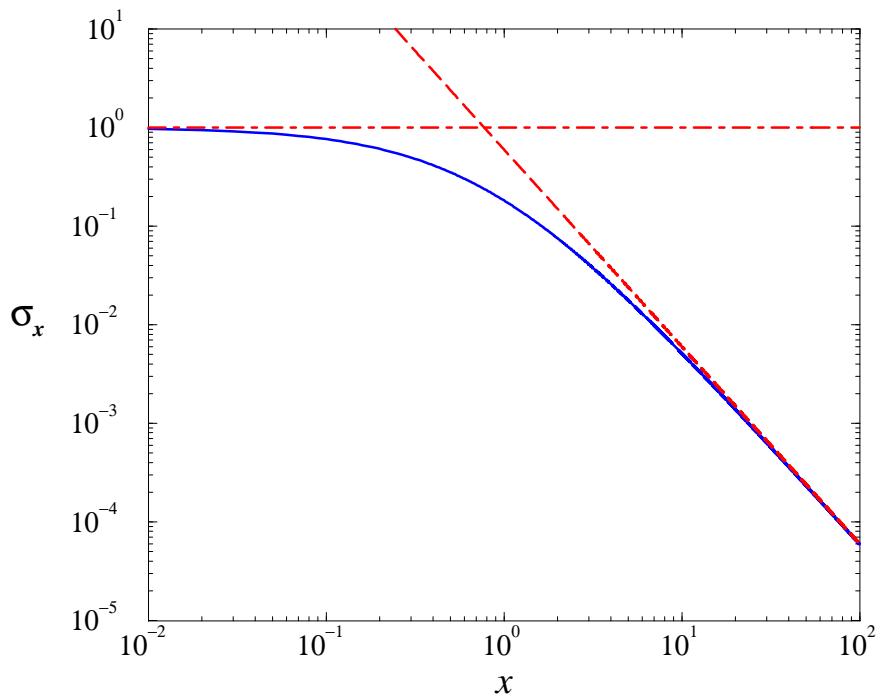}
	\caption{\it Comparison of the various limits for the cross section
	$\sigma(\widetilde{G}\widetilde{G}\rightarrow f\bar{f})$ where
	only the quantity in square brackets $(\equiv\sigma_x)$ in 
	Eq.(\ref{GGffcs}) is plotted. The solid
	curve represents the exact expression for $\sigma_x$, while the
	dot-dashed line depicts $\sigma_x$ in the Equivalence Theorem limit
	$(\protect\sqrt{s}\gg\tilde{m})$ and the dashed line represents
	$\sigma_x$ in the intermediate limit $(\tilde{m}\gg\protect
	\sqrt{s})$.}
\label{fig:GGff}
\end{figure}

The Goldstino annihilation amplitude into fermions is similarly 
obtained from the following Lagrangian:
\begin{eqnarray}
	\label{GGfflag}
	{\cal L} &=& i{2\over\sqrt{3}} {\kappa\over m_{3/2}} (\partial_\mu
	\overline{\widetilde{G}} f_L \partial^\mu \tilde{f}_L - \bar{f}_L
	\partial_\mu\widetilde{G}\partial^\mu \tilde{f}_L^\ast
	+\bar{f}_R\partial_\mu\widetilde{G}\partial^\mu \tilde{f}_R^\ast
	-\partial_\mu\overline{\widetilde{G}} f_R \partial^\mu \tilde{f}_R)
	\nonumber \\
	&+&{1\over 6} {\kappa^2\over m_{3/2}^2}\bar{f} \gamma_5
	\gamma_\sigma f\,\partial^\mu\overline{\widetilde{G}} 
	\gamma_5\gamma^\sigma \partial_\mu\widetilde{G} 
	+i {\kappa\over 4} [\bar{f}(\gamma^\nu\partial^\mu
	+\gamma^\mu\partial^\nu)f -(\partial^\mu\bar{f}\gamma^\nu+
	\partial^\nu\bar{f}\gamma^\mu)f] h_{\mu\nu}
	\nonumber \\
	&-&{1\over 6}~{\kappa\over m_{3/2}} ~(\eta^{\lambda\mu}
	\eta^{\rho\nu}+\eta^{\lambda\nu}\eta^{\rho\mu}-\eta^{\mu\nu}
	\eta^{\lambda\rho})~\partial_\mu\overline{\widetilde{G}}~
	\partial_\nu\widetilde{G}~h_{\lambda\rho}
	\nonumber \\
	&+&{i\over 6}~{\kappa\over m_{3/2}^2}~\epsilon^{\mu\sigma\nu(\lambda}
	~\partial_\mu\overline{\widetilde{G}}~\gamma_5~[\gamma_\sigma,
	\sigma^{\rho)
	\tau}]_+~\partial_\nu\widetilde{G}~\partial_\tau h_{\lambda\rho}
	\nonumber \\
	&-& {\kappa\over 2}\sum_i d_i \bar{f} \gamma^\mu f \partial_\mu P_i
	- {1\over 6} {\kappa\over m_{3/2}^2} \sum_i d_i~\partial_\mu 
	\overline{\widetilde{G}} \gamma^\nu\gamma_5 \partial^\mu 
	\widetilde{G} \partial_\nu P_i 
\end{eqnarray}
where $f$ denotes the fermion Dirac spinor. Note that due to gauge 
invariance there is no fermion coupling to the scalar $S_i$. The relevant 
Feynman diagrams are depicted in Fig.~\ref{fig:FDGGff}. 
Besides the sfermion, graviton and hidden sector
scalar exchange diagrams, there is also a four-Fermi interaction term due 
to the nonrenormalisability of the supergravity Lagrangian. Assuming that 
the fermions are relativistic the cross section is calculated to be
\begin{equation}
	\label{GGffcs}
	\sigma(\widetilde{G}\widetilde{G}\rightarrow f \bar{f})
	={1\over 108\pi} {\kappa^4\over m_{3/2}^4}
	m_{\tilde{f}}^4 s\left[1+{3x(-1+2x+4x^2)\over 1+x} 
	-12x^3 \log (1+{1\over x})\right]
\end{equation}
where $x=m_{\tilde{f}}^2/s$ and the contribution from the pseudoscalar $P_i$
diagram turns out to be zero. In the limit that $\sqrt{s} \gg m_{\tilde{f}}$
one can easily check that (\ref{GGffcs}) agrees with the Equivalence Theorem.
If instead one considers the intermediate limit $m_{\tilde{f}} \gg \sqrt{s}$
then the annihilation cross section (\ref{GGffcs}) becomes
\begin{equation}
	\label{GGffcsil}
	\sigma(\widetilde{G}\widetilde{G}\rightarrow f \bar{f})
	={1\over 180\pi} {\kappa^4\over m_{3/2}^4} s^3
\end{equation}
where again the energy dependence is different than that in the high
energy limit $\sqrt{s}\gg m_{\tilde f}$.
The various limiting behaviours are shown in Fig.~\ref{fig:GGff}. One can 
clearly see in the figure that the Equivalence Theorem limit is not a good 
approximation when $\sqrt{s} \ll \tilde{m}$ and infact the cross section
is considerably smaller.

To obtain the cross section for Goldstino decay into massless neutrinos 
one simply neglects the right handed components in the Lagrangian 
(\ref{GGfflag}). In this case one finds
\begin{equation}
	\label{GGnncs}
	\sigma(\widetilde{G}\widetilde{G}\rightarrow \nu_L \bar{\nu}_L)
	={1\over 2} \sigma(\widetilde{G}\widetilde{G}\rightarrow f \bar{f}).
\end{equation}
The total Goldstino annihilation cross section is given by the sum of
(\ref{GGggcs}) and (\ref{GGffcs}) and will be used to calculate the 
thermally averaged annihilation rate in the early universe.

\section{Thermally averaged annihilation rate}

The annihilation cross sections obtained in the previous section were
calculated at a temperature $T=0$.
In the early universe we need to average over the statistical distributions
of the colliding particles in the thermal heat bath. The thermally averaged
annihilation cross section times velocity for the scattering process
$1+2\rightarrow F$ is given by \cite{gg}
\begin{equation}
	\label{tacs}
	\langle\sigma v_{M\o l}\rangle = {\int dn_1^{eq} dn_2^{eq} \sigma
	v_{M\o l}\over \int dn_1^{eq} dn_2^{eq}}
\end{equation}
where 
\begin{equation}
	\label{dn}
	dn_i^{eq} = f(E_i,t) g_i {d^3 p_i\over (2\pi)^3}
\end{equation}
and $f(E_i,t)$ is the statistical distribution function and $g_i$ is the 
number of internal degrees of freedom for the particle species $i$. 
The factor $v_{M\o l}=\sqrt{(p_1\cdot p_2)^2 -m_1^2 m_2^2}/(E_1 E_2)$ is 
known as the M\o ller velocity (see Ref.~\cite{gg}) and $\sigma$ is the 
sum over all possible annihilation channels of particles 1 and 2. 
We have neglected the Pauli blocking 
factors $1-f$ for final state particles and antiparticles in (\ref{tacs}), 
which typically amounts to a $10\%$ correction in the determination
of the decoupling temperature \cite{enq}.

In the case of Goldstino annihilation we are assuming that 
$E\gg m_{3/2}$ so that the Goldstinos will be relativistic. 
Note that for Goldstinos which obey Fermi-Dirac statistics,
$f(E,t)=1/(e^{E/T}+1)$ the equilibrium number density 
is
\begin{equation}
	\label{numd}
	n_{\widetilde G} = {3\over 2\pi^2} \zeta(3) T^3
\end{equation}
where $\zeta(3)\simeq 1.202$.
In the early universe we will be interested in the role of the Goldstino 
at temperatures $T\sim{\cal O}$(100 MeV). At these times 
$\sqrt{s}\ll \tilde{m}$ and infact $x=10^6$ for $\tilde{m}\sim {\cal O}
(100\,$GeV).
If we parametrise the annihilation cross section
as $\sigma=\sum_i \hat{\sigma}_i s^{n_i}$ where $s$ is the Mandelstam 
variable 
and $\hat{\sigma}_i$ is a constant, then the thermally averaged annihilation
cross section times velocity for the Goldstino in the limit $T\ll \tilde{m}$
is given by
\begin{equation}
	\label{gtacs}
	\langle\sigma v_{M\o l}\rangle = {1\over 9 \zeta(3)^2} \sum_i 
	{2^{2n_i+3}\over (n_i+2)} I(n_i)^2~\hat{\sigma}_i~T^{2n_i}
\end{equation}
where $I(n_i)=\int_0^\infty dy~{y^{n_i+2}\over e^y+1}$, $n_i$ are 
integers and the sum is over all annihilation channels.

Using the expressions (\ref{GGggcsil}) and (\ref{GGffcsil}) 
for the Goldstino annihilation cross sections derived in the previous 
section we obtain
\begin{equation}
	\label{gchan}
	\sigma_A=\sigma(\widetilde{G}\widetilde{G}\rightarrow \gamma\gamma)
	+\sum_f\sigma(\widetilde{G}\widetilde{G}\rightarrow f \bar{f})=
	\hat{\sigma}_{\gamma\gamma} s^2 +\sum_f \hat{\sigma}_{f \bar{f}}s^3,
\end{equation}
where we have summed over all possible fermion pairs in the final state. 
Assuming a photino mass $m_{\tilde\gamma}\sim{\cal O}$(100 GeV), the dominant 
part of the total cross section actually comes from the Goldstino annihilation
into photons so that the thermally averaged cross section times velocity
is approximately
\begin{equation}
	\label{thav}
	\langle\sigma_A v_{M\o l}\rangle \simeq 1800{\zeta(5)^2\over
	\zeta(3)^2}\hat{\sigma}_{\gamma\gamma} T^4,
\end{equation}
where $\zeta(5)\simeq 1.037$.
The average Goldstino annihilation rate which is defined to be 
$\Gamma_A=n_{\widetilde{G}}\langle\sigma_A v_{M\o l} \rangle$ is then 
given by
\begin{equation}
	\label{grate}
	\Gamma_A \simeq {75 \over 16\pi^3} {\zeta(5)^2\over \zeta(3)}
	{\kappa^4\over m_{3/2}^4} m_{\tilde{\gamma}}^2 T^7
	\simeq 0.135{\kappa^4\over m_{3/2}^4} m_{\tilde{\gamma}}^2 T^7
\end{equation}
where $\hat{\sigma}_{\gamma\gamma}$ is determined from Eq. (\ref{GGggcsil}).
This rate will be used to calculate the Goldstino decoupling temperature.

\section{Goldstino decoupling temperature}

During the radiation dominated era of the universe, the energy density which
is dominated by relativistic particles, can be expressed in terms
of the photon energy density $\rho_\gamma(T)$ as
\begin{equation}
	\label{edu}
	\rho(T)={1\over 2}g_{\rho}(T) \rho_\gamma(T)=
	g_{\rho}(T) {\pi^2\over 30} T^4, 
\end{equation}
where the effective number of relativistic degrees of freedom of the bosons 
$(B)$ and fermions $(F)$ present is
\begin{equation}
	\label{effdof}
	g_{\rho}(T) = \sum_B g_B \left({T_B\over T}\right)^4
	+{7\over 8}\sum_F g_F \left({T_F\over T}\right)^4.
\end{equation}
In Eq.(\ref{effdof}) $g_B (g_F)$ is the number of internal degrees of freedom
for each boson (fermion) and $T_{B,F}$ represent the possibility of the 
decoupled particles having a temperature which differs from the photon
temperature, $T$. The Hubble expansion rate during this era will be 
\begin{equation}
	\label{hubble}
	H=\sqrt{{4\pi^3\over 45}} g_\rho^{1/2} \kappa T^2.
\end{equation}
The Goldstinos thermally decouple from the heat bath when their annihilation
rate $\Gamma_A \lesssim H$. Using (\ref{grate}) and (\ref{hubble}) one 
finds that the Goldstino decoupling temperature is 
\begin{equation}
	\label{gdt}
	T_D\simeq {5\over 3}~g_\rho^{1/10}~m_{3/2}^{4/5}~
	m_{\tilde{\gamma}}^{-2/5}~{\kappa^{-3/5}}.
\end{equation}
This equation shows that as the gravitino mass becomes lighter, it causes
the Goldstino to decouple later in the evolution of the universe. The later
the Goldstino decouples, the greater the possibility of the Goldstino 
interfering with nucleosynthesis. Clearly there exists a lower bound on the
gravitino mass for which the predictions of nucleosynthesis are not affected.

During nucleosynthesis the energy density of new massless particles, $i$
is equivalent to an effective number $\Delta N_\nu$ of additional doublet
neutrinos:
\begin{equation}
	\label{effno}
	\Delta N_\nu = f_{B,F} \sum_i {g_i\over 2} \left[{g_\rho(T_\nu)\over
	g_\rho(T_{D_i})}\right]^{4/3}
\end{equation}
where $f_B=8/7$ for bosons, $f_F=1$ for fermions and $g_i$ is the number
of internal degrees of freedom of the particle species $i$ \cite{oss}.
At neutrino decoupling the only known particles which can contribute to 
$g_\rho$ are $\gamma,e^{\pm},\nu_e,\bar{\nu}_e,\nu_\mu,\bar{\nu}_\mu,
\nu_\tau$, and $\bar{\nu}_\tau$. This means that the effective number of 
degrees of freedom at neutrino decoupling are, using Eq.(\ref{effdof})
\begin{equation}
	\label{ndof}
	g_\rho(T_\nu)=2+{7\over 8}(4+3\times2)={43\over 4}.
\end{equation}
Assuming the conservative estimate that $\Delta N_\nu < 0.6$ \cite{olive},
places a lower
bound on the Goldstino decoupling temperature, $T_D$. For example, if 
$T_D=T_\nu \simeq{\cal O}({\rm MeV})$ then according to 
(\ref{effno}), $\Delta N_\nu =1$ and the Goldstino would behave like 
an extra neutrino family. This would mean that $T_D > T_\nu$.
A stronger bound can be obtained by supposing that the Goldstino decouples
during the temperature range $T_\nu< T_D <T_\mu$, where
$T_\mu$ is the muon decoupling temperature. When the Goldstino
decouples the effective number of degrees of freedom would be
\begin{equation}
	\label{dof}
	g_\rho(T_D)=2+{7\over 8}(4+3\times2+2)={25\over 2}.
\end{equation}
Using Eq.(\ref{effno}) this will contribute an amount $\Delta N_\nu =0.82$,
which means that the Goldstino decoupling temperature $T_D > 
T_\mu\simeq {\cal O}(100{\rm MeV})$.
Imposing this condition on the expression (\ref{gdt}) leads to the lower
bound on the gravitino mass
\begin{equation}
	\label{lb}
	m_{3/2} \gtrsim 10^{-6}{\rm eV} \left({m_{\tilde{\gamma}}
	\over 100\,{\rm GeV}}\right)^{1/2}.
\end{equation}
This bound is much weaker than that quoted by Moroi et al \cite{moroi}
because the Goldstino annihilation cross section is not as large
as assumed by those authors.

The mass bound (\ref{lb}) may be strengthened slightly by supposing
that muons are also in thermal equilibrium when the Goldstino decouples. 
In this case one obtains $g_\rho(T_D)=16$ and $\Delta N_\nu =0.59$. 
Assuming that the primordial $^4$He abundance rules this contribution
out as well, 
causes the the lower bound on the Goldstino decoupling temperature to 
increase up to the pion mass $T\sim m_\pi$ and
the lower bound (\ref{lb}) to increase slightly.
Clearly, the higher the Goldstino decoupling temperature becomes the 
less it will contribute to $N_\nu$ as many more particles and resonances
contribute to $g_\rho(T_D)$. In order to significantly increase the bound
in future a more accurate estimate of $\Delta N_\nu$ would be needed.

Of course it is likely that there are other beyond the standard model 
particles which could contribute to $\Delta N_\nu$. In this case it may
be more difficult to accommodate the Goldstino, which would cause the bound 
(\ref{lb}) to increase further. This could happen for example if neutrinos 
have Dirac masses. It is known that neutrino Dirac masses can 
contribute a large amount to $\Delta N_\nu$ \cite{dolgov}. Accounting for 
the Goldstino and neutrino masses could place further constraints on the
mass parameters.

\section{Conclusion}

If a spontaneously broken N=1 supergravity theory produces a sufficiently
light gravitino then it will have consequences during the nucleosynthesis
era of the early universe. This may be possible in no-scale supergravity
theories or in theories of low energy dynamical supersymmetry 
breaking with a gauge-mediated messenger sector. The gravitino will obtain
a mass via the superHiggs effect by absorbing the Goldstino.
The enhanced coupling of the Goldstino causes it to interact more strongly
with chiral and vector supermultiplets, which means that the Goldstino
can decouple just prior to the nucleosynthesis era. 

The primordial $^4$He abundance critically depends on the number of massless
neutrino families. If we require that the Goldstino not contribute
significantly to the number of massless neutrino families, a lower bound 
on the gravitino mass can be obtained. Previous lower bounds have ranged 
from $10^{-4}-10^{-2}$eV \cite{fayet,moroi}. 
By calculating the Goldstino annihilation cross section into leptons and 
photons in the limit $\sqrt{s}\ll \tilde{m}$ we were able to show that this
bound is considerably weaker than previous estimates. If $N_\nu < 3.6$ then
typically $m_{3/2} \gtrsim 10^{-6}$eV for $m_{\tilde{\gamma}}\simeq
{\cal O}$(100 GeV). This bound complements previous gravitino mass 
bounds derived from collider experiments \cite{fayetcoll,dicus} and 
astrophysics \cite{fs}. 

In addition a bound on the supersymmetry breaking scale, $\Lambda$ can 
also be obtained for various scenarios of supersymmetry breaking dynamics.
Assuming that $m_{3/2} \simeq \Lambda^2/M_{pl}$, the bound on the
gravitino mass implies that the scale of supersymmetry breaking, 
$\Lambda \gtrsim 100$GeV. In particular this would set a lower bound 
(100 GeV) on the scale of the supercolour sector in the recent gauge 
mediated models. These bounds are right at the forefront of existing 
collider energy scales.

\bigskip

\section{Acknowledgements}

We would like to thank Ganesh Devaraj, Marty Einhorn and Steve Martin for 
useful conversations. 
This work was supported by the U.S. Department of Energy at the University 
of Michigan.

\vfil\eject

\end{document}